%% file: main.tex
\documentclass{sig-alternate-05-2015}

\usepackage[utf8]{inputenc}
\usepackage{amsmath}
\usepackage{amsfonts}
\usepackage{amssymb}
\usepackage{algorithm}
\usepackage{algpseudocode}
\usepackage{lmodern}
\usepackage{mathptmx}
\usepackage[skip=8pt,font=normalsize]{caption}
\usepackage{epigraph}
\usepackage{url}
\usepackage[hidelinks]{hyperref}
\usepackage[usenames,dvipsnames,svgnames,table]{xcolor} 
\usepackage{graphicx}
\usepackage{booktabs}
\usepackage{adjustbox}
\usepackage{array}
\usepackage{xstring}
\usepackage{threeparttable}
\usepackage{microtype}
\SetExtraKerning{encoding=*,font=*}{\textemdash={120,120}} 
\usepackage{balance}
\usepackage{multirow}
\usepackage[toc,page]{appendix}
\usepackage{enumitem}

\usepackage{soul}
\usepackage{xspace}

\newcommand{\etal}{{\it et al.}}

\usepackage{wasysym}
\input{Macros}

\renewcommand{\paragraph}[1]{\smallskip\noindent\textbf{#1}\quad}

\def\msr{$^\dagger$}
\def\umich{$^\triangleleft$}
\def\mit{$^\diamond$}
\def\lux{$^\circ$}
\def\cal{$^\ddagger$}
\def\melb{$^\triangleright$}
\def\gwu{$^\S$}
\def\rice{$^\star$}

\author{
    {\rm  \large  Matthew Bernhard\umich\quad 
Josh Benaloh\msr\quad
J. Alex Halderman\umich\quad 
Ronald L. Rivest\mit} \\
{\rm \large Peter Y. A. Ryan\lux\quad
Philip B. Stark\cal\quad 
Vanessa Teague\melb\quad 
Poorvi L. Vora\gwu\quad 
Dan S. Wallach\rice }\vspace{.4cm} \\
\normalsize
\msr Microsoft Research\quad
\umich University of Michigan\quad
\mit Massachusetts Institute of Technology\quad
\lux  University of Luxembourg\\
\normalsize
\cal  University of California at Berkeley\quad
\melb  University of Melbourne\quad
\gwu  George Washington University\quad
\rice  Rice University
}

\title{Public Evidence from Secret Ballots}

\begin{document}
\pagenumbering{arabic}
\thispagestyle{empty}

\makeatletter\def\@copyrightspace{\relax}\makeatother

\maketitle

\input{Abstract}
\input{Introduction}
\input{Requirements}

\input{TechTools}

\input{Designs}

\input{InternetVoting}

\input{ALookAhead}

\section*{Acknowledgments}

This work was supported in part by the U.S. National Science Foundation awards
CNS-1345254, CNS-1409505, CNS-1518888, CNS-1409401, CNS-1314492, and 1421373, the Center for Science of
Information STC (CSoI), an NSF Science and Technology Center, under grant
agreement CCF-0939370, the Maryland Procurement Office under contract
H98230-14-C-0127, and FNR Luxembourg under the PETRVS Mobility grant.  

\makeatletter
\def\thebibliography#1{%
  \ifnum\addauflag=0\addauthorsection\global\addauflag=1\fi
  \section[References]{
    {References} 
    {\vskip -3pt plus 1.55pt } 
    \@mkboth{{\refname}}{{\refname}}%
  }%
  \list{[\arabic{enumi}]}{%
    \settowidth\labelwidth{[#1]}%
    \leftmargin\labelwidth
    \advance\leftmargin\labelsep
    \advance\leftmargin\bibindent
    \parsep=0pt\itemsep=1pt 
    \itemindent -\bibindent
    \listparindent \itemindent
    \usecounter{enumi}
  }%
  \let\newblock\@empty
  \raggedright 
  \sloppy
  \sfcode`\.=1000\relax
}
\makeatother

{\clubpenalty=0\widowpenalty=0\balance\small
\bibliographystyle{abbrv}
\bibliography{e-vote}}



\end{document}

%% file: Macros.tex
\newcommand{\PaV}{Pr\^et \`a Voter}

\newdef{definition}{\vspace{-4pt}Definition\vspace{-4pt}}

\newcommand{\OpenProblem}[1]{
\vspace{6pt}
\noindent\fbox{\parbox{0.97\columnwidth}{{\bf Open problem:} #1}}
\vspace{4pt}
}

\newcommand{\OpenProblems}[1]{
\vspace{6pt}
\noindent\fbox{\parbox{0.97\columnwidth}{{\bf Open problems:} #1}}
\vspace{4pt}
}

\newcommand{\StL}{Sainte-Lagu\"e}

\let\OLDitemize\itemize
\renewcommand\itemize{\OLDitemize\setlength{\itemsep}{.5pt}}
\setlist[itemize]{leftmargin=5.5mm,rightmargin=1mm}

%% file: Abstract.tex
\begin{abstract}

Elections seem simple---\emph{aren't they just counting?}
But they have a unique, challenging combination of security and privacy requirements.
The stakes are high; the context is adversarial; the electorate needs to be convinced
that the results are correct; and the secrecy of the ballot must be ensured.
And they have practical constraints:
time is of the essence, and
voting systems need to be affordable and maintainable, and usable
by voters, election officials, and pollworkers.
It is thus not surprising that voting is a rich research area
spanning theory, applied cryptography, practical systems analysis,
usable security, and statistics.
Election integrity involves two key concepts: 
\emph{convincing evidence that outcomes are correct} and
\emph{privacy}, which amounts to \emph{convincing assurance that there is no evidence}
about how any given person voted. 
These are obviously in tension.
We examine how current systems walk this tightrope.

\end{abstract}

%% file: Introduction.tex
\section{Introduction: What is the evidence?}
\label{sec:introduction}


\epigraph{The Russians did three things $\ldots$ The third is that they tried, and they were not successful, but they still tried, to get access to voting machines and vote counting software, to play with the results}{\textit{Former CIA Acting Director Michael Morell}, Mar.~15, 2017}

\epigraph{These are baseless allegations substantiated with nothing, done on a rather amateurish, emotional level}{\textit{Kremlin spokesman Dmitry Peskov}, Jan.~9, 2017}

\epigraph{It would take an army to hack into our voting system. }{\textit{Tom Hicks, EAC Commissioner}, Oct.~6, 2016}





It is not enough for an election to produce the correct outcome.
The electorate must also be convinced that the announced result reflects
the will of the people.
And for a rational person to be convinced requires evidence.

Modern technology---computer and communications systems---is 
fragile and vulnerable to programming errors and undetectable manipulation.
No current system that relies on electronic technology alone to capture and
tally votes can provide
convincing evidence that election results are accurate without endangering or sacrificing
the anonymity of votes.\footnote{%
    Moreover, the systems that come closest are not readily usable by a typical voter.
}

Paper ballots, on the other hand, have some very helpful security properties:
they are readable (and countable, and re-countable) by humans; they are relatively durable;
and they are tamper-evident.
Votes cast on paper can be counted using electronic technology; then the accuracy of the count can be
checked manually to ensure that the technology functioned adequately well.
Statistical methods allow the accuracy of the count to be assessed by examining
only a fraction of the ballots manually, often a very small fraction.
If there is also convincing evidence that the collection of ballots has been
conserved (no ballots added, lost, or modified) then this combination---voter-verifiable 
paper ballots, a mechanized count, and a manual check of the
accuracy of that count---can provide convincing evidence that announced
electoral outcomes are correct.

Conversely, absent convincing evidence that the paper trail has been conserved,
a manual double-check of electronic results against the paper trail will not be convincing.
If the paper trail has been conserved adequately, then a full manual tally
of the ballots can correct the electronic count if the electronic count is incorrect.

These considerations have led many election integrity advocates to push for a voter-verifiable
paper trail (VVPAT).\footnote{%
  Voter-marked paper ballots or ballots marked using a ballot-marking device are preferable
  to VVPAT, a cash-register style printout that the voter cannot touch.
}

In the 2016 presidential election, about three quarters of Americans voted using 
systems that generated voter-verifiable paper records.
The aftermath of the election proved that even if 100\% of voters had used
such systems, it would not have sufficed to provide convincing evidence that the
reported results are accurate.
\begin{itemize}
    \setlength\itemsep{.5pt}
   \item No state has (or had) adequate laws or regulations to ensure that the paper trail is conserved adequately,
 and that provide evidence to that effect.
   \item No state had laws or regulations that provided adequate manual scrutiny of the paper
 to ensure that the electronically generated results are correct; most still do not.
   \item  Many states that have a paper trail also have laws that make it hard for anyone to 
check the results using the paper trail---even candidates with war chests for litigation.
Not only can other candidates fight attempts to check the results, the states themselves can
fight such attempts. 
This treats the paper as a nuisance, rather than a safeguard.
\end{itemize}

The bottom line is that the paper trail is not worth the paper it's printed on.
Clearly this must change.

Other techniques like \emph{software independence} and \emph{end-to-end verifiability} can
offer far greater assurance in the accuracy of an election's outcome, but these methods
have not been broadly applied.


\subsection{Why so hard?}
Several factors make it difficult to generate convincing evidence that reported
results are correct.  The first is the trust model.

\paragraph{No one is trusted}
In any significant election, voters, election officials, and equipment and
software 
cannot necessarily be
trusted by anyone with a stake in the outcome.  Voters, operators, system
designers, manufacturers, and external parties are all potential
adversaries.

\paragraph{\bf The need for evidence}
Because officials and equipment may not be trustworthy, elections should be
\emph{evidence-based}.  Any observer should be able to verify the reported
results based on trustworthy evidence from the voting system.  Many in-person voting
systems fail to provide sufficient evidence; and as we shall see Internet
systems scarcely provide any at all.

\paragraph{\bf The secret ballot}
Perhaps the most distinctive element of elections is the \emph{secret ballot},
a critical safeguard that defends against vote selling and voter
coercion.  In practical terms, voters should not be able to prove how they
voted to anyone, \emph{even if they wish to do so}. 
This restricts
the types of evidence that can be produced by the voting system. 
Encryption alone is not sufficient, since the voters may choose to
reveal their selections in response to bribery or coercion. 

\smallskip The challenge of voting is thus to use fragile technology to produce
trustworthy, convincing \emph{evidence} of the correctness of the outcome while protecting
voter \emph{privacy} in a world \emph{where no person or machine may be
trusted}.  
The resulting voting system and its security features must also be
\emph{usable} by regular voters.

The aim of this paper is to explain the important requirements
of secure elections and the solutions already available from e-voting research, then to 
identify the most important directions for research. 

Prior to delving into our discussion, we need to make a distinction in
terminology.  
\emph{Pollsite} voting systems are those in which voters record
and cast ballots at predetermined locations, often in public areas with strict
monitoring. \emph{Remote} voting refers to a system where voters fill out ballots anywhere, and
then send them to a central location to cast them, either physically mailing
them in the case of vote-by-mail, or sending them over the Internet in the case
of Internet voting. 

The next section defines the requirements, beginning with notions of election evidence,
then considering privacy, and concluding with more general usability and security requirements.
Section~\ref{sec:Tools} describes the cryptographic, statistical, and
engineering tools that have been developed for designing voting systems
with verifiably correct election outcomes.
Section~\ref{sec:Designs} discusses the challenge of satisfying
our requirements for security using the tools presented in real-world 
election systems. 
Section~\ref{sec:InternetVoting} concludes with the promise and problems
associated with Internet voting.

%% file: Requirements.tex
\section{Requirements for Secure Voting}
\label{sec:Requirements}

\epigraph{Trustworthiness before trust}{\textit{Onora O'Neill}}

\input{TrustVerifyEvidence}


\input{Authentication}

\input{PrivacyAndCoercion}

\input{DoS}

\input{OtherRequirements}  

%% file: TrustVerifyEvidence.tex
\vspace{-10pt}
\subsection{Trust, Verifiability, and Evidence}
\label{sec:trust}
For an election to be accepted as legitimate, the outcome should be
convincing to all---and in particular to the losers---leaving no valid
grounds to challenge the outcome.  Whether elections are conducted
by counting paper ballots by hand or
using computer technology, the possibility of error or fraud necessitates
assurances of the accuracy of the outcome.




 It is clear that a naive introduction of computers into voting
 introduces the possibility of wholesale and largely undetectable
 fraud.  If we can't detect it, how can we prevent it?

\subsubsection{Risk-Limiting Audits}
Statistical post-election audits provide assurance that a reported outcome is
correct, by examining some or all of an \emph{audit trail} consisting of
durable, tamper-evident, voter-verifiable records. Typically the audit trail
consists of paper ballots.

The \emph{outcome} of an election is the set of winners.
An outcome is incorrect if it differs from the set of winners output by a perfectly accurate
manual tabulation of the audit trail.

\vspace{-4pt}
\begin{definition}
An audit of an election contest is a \emph{{\bf risk-limiting audit (RLA)} with risk limit $\alpha$}
if it has the following two properties:
\begin{enumerate}
\item If the reported contest outcome under audit is incorrect,
the probability that the audit leads to correcting the outcome is at least $1-\alpha$.
\item The audit never indicates a need to alter a reported outcome that is correct.
\end{enumerate}
\end{definition}
\vspace{-4pt}
(In this context, ``correct'' means ``what a full manual tally of the paper
trail would show.'' If the paper trail is unreliable, a RLA in general cannot
detect that.  RLAs should be preceded by ``compliance audits'' that check
whether the audit trail itself is adequately reliable to determine who won.)
Together, these two properties imply that post-RLA, either the reported set of
winners is the set that a perfectly accurate hand count of the audit trail
would show, or an event with probability no larger than $\alpha$ has occurred.
(That event is that the outcome was incorrect, but the RLA did not lead to
correcting the outcome.) RLAs amount to a limited form of probabilistic error
correction: by relying on appropriate random sampling of the audit trail and
hypothesis tests, they have a known minimum probability of correcting the
outcome.  They are not designed to ensure that the reported numerical tally is
correct, only that the outcome is correct.

The following procedure is a trivial RLA: with probability $1-\alpha$, perform
a full manual tally of the audit trail. Amend the outcome to match the set of
winners the full hand count shows if that set is different.

The art in constructing RLAs consists of maintaining the risk limit while
performing {\em less work} than a full hand count when the outcome is correct.
Typically, this involves framing the audit as a sequential test of the
statistical hypothesis that the outcome is incorrect. To reject that hypothesis
is to conclude that the outcome is correct. RLAs have been developed for
majority contests, plurality contests, and vote-for-$k$ contests and complex
social choice functions including D'Hondt and other proportional representation
rules---see below.  RLAs have also been devised to check more than one election
contest simultaneously~\cite{stark-simple}.

\subsubsection{Software Independence}
\label{sec:software-independence}


Rivest and Wack
introduced a definition targeted specifically at
detecting misbehavior in computer-based elections:

\vspace{-4pt}
\begin{definition}\cite{rivest2008notion}
A voting system is {\bf software independent} if an undetected change
or error in its software cannot cause an undetectable change or error
in an election outcome.
\end{definition}
\vspace{-4pt}

 Software independence clearly expresses that it should not be necessary
 to trust software to determine election outcomes, 
 but it does not say what procedures or types of evidence should be trusted instead.
 A system that is not software
 independent \emph{cannot} produce a convincing evidence trail, but neither
 can a paper-based system that does not ensure that the paper trail is complete
 and intact, a cryptographic voting system that relies on an invalid
 cryptographic assumption, or a system that relies on audit procedures
 but lacks a means of assuring that those procedures are properly followed.  
 We could likewise demand independence of many other kinds
 of trust assumptions: hardware, paper chain-of-custody, cryptographic setup,
 computational hardness, procedures, good randomness generation {\it etc.}

Rivest and Wack also define a stronger form of the property that
includes error recovery:

\begin{definition}\cite{rivest2008notion}
A voting system is {\bf strongly software independent} if it is
software independent and a detected change or error in an
election outcome (due to the software) can be
corrected without rerunning the election.
\end{definition}

A strongly software-independent system can recover from software
errors or bugs, but that recovery in turn
is generally based on some other trail of evidence.

A software independent system can be viewed as a form of tamper-evident system:
a material software problem leaves a detectable trace.
\emph{Strongly} software independent systems are resilient: not only do 
material software problems
leave a trace, the overall election system can recover from a detected problem.

One mechanism to provide software independence is to record votes on a paper
record that provides physical evidence of voter's intent, can be inspected
by the voter prior to casting the vote, and---if preserved intact---can later be
manually audited to check the election outcome. 
Risk-limiting audits (see Section~\ref{sec:Statistics}) can then achieve a pre-specified level of
assurance that results are correct; machine assisted risk-limiting audits~\cite{machineRLA}, 
can help minimize the amount of labor required for legacy systems that do not
provide a cast-vote record for every ballot, linked to the corresponding ballot.


\OpenProblems{
\begin{itemize}
\item How can systems handle errors in the event that elections don't verify? Can they recover?
\end{itemize}
}
\subsubsection{End-to-end verifiability}
\label{sec:E2E}

The concern regarding fraud and desire for transparency has motivated the
security and crypto communities to develop another approach to voting system
assurance: \emph{end-to-end verifiability} (E2E-V). An election that is
end-to-end verifiable achieves software independence together with the
analagous notion of hardware independence as well as independence from actions
of election personnel and vendors.  Rather than attempting to verify thousands
of lines of code or closely monitor all of the many processes in an election,
E2E-V focuses on providing a means to detect errors or fraud in the process of
voting and counting. 
The idea behind E2E-V is to enable voters
themselves to monitor the integrity of the election; democracy for the people
by the people, as it were.  
This is challenging because total transparency is
not possible without undermining the secret ballot, hence the mechanisms to
generate such evidence have to be carefully designed.





\vspace{-3pt}
\begin{definition} {\it (adapted from \cite{benaloh2015end})}
A voting system is {\bf end-to-end verifiable} if it has the following three kinds of verifiability:
\begin{itemize}
    \setlength\itemsep{.5pt}
\item {\bf Cast as intended:}\enspace Voters can independently verify that their selections
  are correctly recorded.
\item {\bf Collected as cast:}\enspace Voters can independently verify that the representation of their vote is
correctly collected in the tally.
\item {\bf Tallied as collected:}\enspace Anyone can verify that every
  well-formed, collected vote is correctly included in the tally.
\end{itemize}
If verification relies on trusting entities, software, or hardware, the voter
and/or auditor should be able to choose them freely.  Trusted procedures, if
there are any, must be open to meaningful observation by \emph{every} voter.
\end{definition}

Note that the above definition allows each voter to check that her vote is
correctly collected, thus ensuring that attempts to change or delete cast votes
are detected.  In addition, it should also be possible to check the list of
voters who cast ballots, to ensure that votes are not added to the collection
({\it i.e.,} to prevent ballot-box stuffing).  This is called \emph{eligibility
verifiability} \cite{kremer2010election,smyth2010towards}.
\looseness=-1

\input{CollectionAccountable}
\subsubsection{Dispute Resolution}
\label{sec:dispute}

While accountability helps secure the election process, it is
not very useful if there is no way to handle disputes. 
If a voter claims, on the basis of accountability checks provided by a system,
that something has gone wrong, there needs to be a mechanism to address this. 
This is known as \emph{dispute resolution}:

\begin{definition} \cite{DBLP:conf/voteid/KaczmarekWCFRRVZ13}
    A voting system is said to have {\bf dispute resolution} if, when there is a
    dispute between two participants regarding honest participation, a third
    party can correctly resolve the dispute.
\end{definition}

An alternative to dispute resolution is dispute-freeness:

\begin{definition}\cite{Kiayias:2002:SEP:648119.746910}
    A {\bf dispute-free} voting system has built-in prevention mechanisms that
    eliminate disputes among the active participants; any third party can
    check whether an active participant has cheated.
\end{definition}

\OpenProblems {
\begin{itemize}
    \setlength\itemsep{.5pt}
    \item Can effective dispute resolution for all classes of possible errors exist
          in a given system?
    \item Are there other reasonable definitions and mechanisms for dispute resolution?
    \item Can a system offer complete dispute resolution capabilities in which every
dispute can be adjudicated using evidence produced by the election system?
\end{itemize}
}

\subsubsection{From Verifiable to Verified}

Constructing a voting system that creates sufficient evidence
to reveal problems is not enough on its own.  
That evidence must
actually be used---and used appropriately---to ensure the
accuracy of election outcomes.

An election result may not be verified, even if it is 
generated by an end-to-end verifiable voting system.
For verification of the result,
 we need several further conditions to be satisfied:
\begin{itemize}
    \setlength\itemsep{.5pt}
\item Enough voters and observers must be sufficiently diligent in performing
  the appropriate checks.
\item Random audits (including those initiated by voters) must be sufficiently
    extensive and unpredictable that changes that affect election outcomes
    have a high chance of being detected.
\item If checks fail, this must be reported to the authorities who, in
  turn, must take appropriate action.
\end{itemize}
These issues involve complex human factors, including voters'
incentives to participate in verification.  
Little work has been done on this aspect of the problem.

An E2E-V system might give an individual voter assurance that her vote
has not been tampered with \emph{if} that voter performs certain
checks.  However, sufficiently many voters must do this in order to
provide evidence that the election outcome as a whole is correct.
Combining risk-limiting audits with E2E-V systems can provide a
valuable layer of protection in the case that an insufficient number
of voters participate in verification.


Finally, another critical verification problem that has received little
attention to date is how to make schemes that are recoverable in the face of
errors. We do not want to have to abort and rerun an election every time a
check a fails.  
Certain levels of detected errors can be shown to be highly
unlikely if the outcome is incorrect, and hence can be tolerated. 
Other types and patterns of error cast doubt on the outcome
and may require either full inspection or retabulation of the paper trail or, if the paper trail
cannot be relied upon, a new election. 

Both K{\"u}sters \etal~~\cite{kusters2011verifiability} and Kiayias
\etal~\cite{kiayias2015end} model voter-initiated
auditing~\cite{benaloh2007ballot} and its implications for detection of an
incorrect election result.  Both definitions turn uncertainty about voter
initiated auditing into a bound on the probability of detecting deviations of
the announced election result from the truth.



\OpenProblems{
\begin{itemize}
\setlength\itemsep{.5pt}
\item Can systems be designed so that the extent and diligence of
checks performed can be measured?
\item Can verification checks be abstracted from voters, either by embedding
      them in election processes or automating them?
\end{itemize}
}


%% file: CollectionAccountable.tex
\subsubsection{Collection Accountability}


In an E2E-V election protocol, voters can check whether their
votes have been properly counted, but if they discover a problem, there
may not be adequate evidence to correct it.  
An election
system that is \emph{collection-accountable} provides voters with evidence of any
failure to collect their votes. 

\begin{definition} An election system is {\bf collection
    accountable} if any voter who detects that her vote has not been collected has, as
    part of the vote-casting protocol, convincing evidence that can be
    presented to an independent party to demonstrate that the vote has not been
    collected.
\end{definition}

Another form of evidence involves providing each voter with a
code representing her votes, such that knowledge of a correct code is
evidence of casting a particular vote \cite{chaum08:scantegrity}.
Yet another mechanism is a suitable paper receipt.
Forensic analysis may provide evidence that this receipt was not forged by a
voter \cite{VeriScan,starvote-jets}.



\OpenProblems{
\begin{itemize}

\item Can independently verifiable evidence be provided by the voting system for incorrect ballot casting?

\end{itemize}
}

%% file: Authentication.tex
\vspace{-4pt}
\subsection{Voter Authentication}
\label{sec:auth}

A significant challenge for election systems is the credentialing of voters to
ensure that all eligible voters, and no one else, can cast votes. 
This presents numerous questions: what kinds of
credentials should be used?  How should they be issued?  Can they be revoked or
de-activated?  Are credentials good for a single election or for an extended
period?  How difficult are they to share, transfer, steal, or forge?  Can the
ability to create genuine-looking forgeries help prevent coercion?  These
questions must be answered carefully, and until they are satisfied for remote
voting, pollsite voting is the only robust way to address these questions---and
even then, in-person credentialing is subject to forgery, distribution, and
revocation concerns (for instance, the Dominican Republic recently held a
pollsite election where voters openly sold their credentials~\cite{dominican}).
In the U.S., there is concern that requiring in-person credentialing, in the
form of voter ID, disenfranchises legitimate voters.

\OpenProblems{
    \begin{itemize}
        \setlength\itemsep{.5pt}
    \item Is there a sufficiently secure way credential Internet voting?
    \item Can a traditional PKI ensure eligibility for remote voting?
    \item How does use of a PKI change coercion assumptions?
    \end{itemize}
}

%% file: PrivacyAndCoercion.tex
\vspace{-3pt}
\subsection{Privacy, Receipt Freeness, and Coercion Resistance}
\label{sec:PrivacyAndCoercion}


In most security applications, privacy and confidentiality are synonymous. In
elections, however, privacy has numerous components that go well beyond typical
confidentiality.  Individual privacy can be compromised by ``normal'' election 
processes such as a unanimous result.  Voters may be coerced if they can 
produce a proof of how they voted, even if they have to work to do so.

Privacy for votes is a means to an end: if voters don't express their true preferences
then the election may not produce the right outcome.  
This section gives an overview of increasingly strong definitions of what it means for
voters to be free of coercion. 

\subsubsection{Basic Confidentiality}
\label{subsec:confidentiality}

We will take \emph{ballot privacy} to mean that the
election does not leak any information about how any voter voted
beyond what can be deduced from the announced results. 
Confidentiality is not the only privacy requirement in elections, but even
simple confidentiality poses significant challenges.  It is remarkable
how many deployed e-voting systems have been shown to lack even the
most basic confidentiality properties
(e.g.,~\cite{halderman2015new,feldman07diebold,Norway-randomness-bug,California-TTBR-site,EVERESTmain}).


Perhaps more discouraging to basic privacy is the fact that remote voting
systems (both paper and electronic) inherently allow voters to eschew confidentiality. Because
remote systems enable voters to fill out their
ballots outside a controlled environment, anyone can watch over the voter's
shoulder while she fills out her ballot.

In an election---unlike, say, in a financial transaction---even the
candidate receiving an encrypted vote should not be able to decrypt it.
Instead, an encrypted (or otherwise shrouded) vote must remain confidential to
keep votes from being directly visible to election authorities.



Some systems, such as code voting~\cite{cha01:e-vote} and the Norwegian and
Swiss Internet voting schemes, defend privacy against an attacker who controls
the computer used for voting; however, this relies on assumptions about the
privacy and integrity of the code sheet.  Some schemes, such as
JCJ/Civitas~\cite{juels05:e-vote}, obscure who has voted while providing a
proof that only eligible votes were included in the tally.

Several works~\cite{delaune2010verifying}~\cite{kusters2011verifiability},
following Benaloh~\cite{Benaloh:1987:VSE:914093} formalize the notion of
privacy as preventing an attacker from noticing when two parties swap their
votes. 

\OpenProblems{
\begin{itemize}
\item Can we develop more effective, verifiable forms of assurance that vote privacy is preserved?
\item Can we build means of privacy for remote voting through computer-based systems?
\end{itemize}
}

\subsubsection{Everlasting Privacy}

Moran and Naor expressed concern over what might happen to encrypted
votes that can still be linked to their voter's name some decades into
the future, and hence decrypted by superior technology.  They define a
requirement to prevent this:

\begin{definition}\cite{Moran-2006-receipt}
  A voting scheme has {\bf everlasting privacy} if its privacy does
  not depend on assumptions of cryptographic hardness.
\end{definition}


Their solution uses perfectly hiding commitments to
the votes, which are aggregated homomorphically. 
Instead of privacy depending upon a cryptographic
hardness assumption, it is the integrity of an election that depends upon a
hardness assumption; and only a real-time compromise of the assumption can have
an impact.


\subsubsection{Systemic Privacy Loss}
\label{subsec:systemic-privacy}


We generally accept that without further information, a voter is more likely to
have voted for a candidate who has received more votes, but additional data is
commonly released which can further erode voter privacy.  Even if we exclude
privacy compromises, there are other privacy risks which must be managed.  If
voters achieve privacy by encrypting their selections, the holders of
decryption keys can view their votes.  If voters make their selections on
devices out of their immediate control ({\it e.g.} official election
equipment), then it is difficult to assure them that these devices are not
retaining information that could later compromise their privacy.  If voters
make their selections on their own devices, then there is an
even greater risk that these devices could be infected with malware that
records (and perhaps even alters) their selections (see, for instance, the
Estonian system~\cite{estonia}).

\OpenProblems{
\begin{itemize}
    \setlength\itemsep{.5pt} 
\item Are there ways to quantify systemic privacy loss?
\item Can elections minimize privacy loss?
\item Can elections provide verifiable integrity while minimizing privacy loss?
\end{itemize}
}

\vspace{-2pt}
\subsubsection{Receipt-freeness}
\label{subsec:ReceiptFreeness}

Preventing coercion and vote-selling was considered solved with the
introduction of the {\em Australian\/} ballot.  The process of voting
privately within a public environment where privacy can be monitored and
enforced prevents improper influence. Recent systems have complicated this
notion, however. If a voting protocol provides a receipt but is not carefully
designed, the receipt can be a channel for information to the coercive
adversary. 

Benaloh and Tuinstra \cite{Benaloh-1994-receipt} pointed out that passive
privacy is insufficient for resisting coercion in elections:  

\begin{definition}
    A voting system is {\bf receipt free} if a voter is
    unable to prove how she voted \emph{even if she actively colludes with a
    coercer and deviates from the protocol in order to try to
    produce a proof}. 
\end{definition}

Traditional elections may fail receipt-freeness too.  In general, if a vote
consists of a long list of choices, the number of possible votes may be much
larger than the number of likely voters.  This is sometimes called (a failure
of) the {\em short ballot assumption}~\cite{rivest2007three}.  Prior to each
election, coercers assign a particular voting pattern to each voter.  When the
individual votes are made public, any voter who did not cast their pattern can
then be found out.  This is sometimes called the {\em Italian attack\/}, after
a once prevalent practice in Sicily.  It can be easily mitigated when a vote
can be broken up, but is difficult to mitigate in systems like IRV in which the
vote is complex but must be kept together.
Mitigations are discussed in Sections~\ref{sec:complex-audits}
and~\ref{sec:cryptographic-solutions}.

\emph{Incoercibility} has been defined and examined in the universally
composable framework in the context of general multiparty computation
\cite{canetti1996incoercible,unruh2010universally}.  These definitions sidestep
the question of whether the voting function itself allows coercion (by
publishing individual complex ballots, or by revealing a unanimous result for
example)---they examine whether the protocol introduces additional
opportunities for coercion.  With some exceptions (such as
\cite{alwen2015incoercible}), they usually focus on a passive notion of
receipt-freeness, which is not strong enough for voting. 


\subsubsection{Coercion Resistance}
\label{sec:CR}
Schemes can be receipt-free, but not entirely resistant to coercion.  Schemes
like \PaV{} \cite{RyanBHSX09} that rely on randomization for receipt-freeness
can be susceptible to {\em forced randomization}, where a coercer forces a
voter to always choose the first choice on the ballot. Due to randomized
candidate order, the resulting vote will be randomly distributed. If a specific
group of voters are coerced in this way, it can have a disproportionate impact
on the election outcome.

If voting rolls are public and voting is not mandatory,
this has an effect equivalent to prevent {\em forced abstention}, wherein a
coercer refuses to let a voter vote. Schemes that rely on credentialing are
also susceptible to coercion by {\em forced surrender of credentials}. 

One way to fully resist forced abstention is to obscure who voted.  However,
this is difficult to reconcile with the opportunity to verify that only
eligible voters have voted (eligibility verifiability),
though some schemes achieve both \cite{haenni2011secure}.  

Moran and Naor \cite{Moran-2006-receipt} provide a strong definition of receipt
freeness in which a voter may deviate actively from the protocol in order to
convince a coercer that she obeyed.  Their model accommodates forced
randomization.  A scheme is resistant to coercion if the voter can always
pretend to have obeyed while actually voting as she likes.

\vspace{-1pt}
\begin{definition} 
    A voting scheme is {\bf coercion resistant} if there exists a way
    for a coerced voter to cast her vote such that her coercer cannot
    distinguish whether or not she followed the coercer's instructions.
\end{definition}

\vspace{-3pt}

Coercion resistance is defined in \cite{juels05:e-vote} to include receipt
freeness and defence against forced-randomization, forced abstention and the
forced surrender of credentials.  More general definitions include
\cite{kusters2012game}, which incorporates all these attacks along with Moran
and Naor's notion of a coercion resistance strategy.

Note that if the coercer can monitor the voter throughout the vote casting
period, then resistance is futile. For in-person voting, we assume that the
voter is isolated from any coercer while she is in the booth (although this is
questionable in the era of mobile phones).  For remote voting, we need to
assume that voters will have some time when they can interact with the voting
system (or the credential-granting system) unobserved.

\subsubsection{More Coercion Considerations}
\label{subsec:OtherCoercion}

Some authors have tried to provide some protection against coercion without
achieving full coercion resistance.  {\it Caveat
	coercitor}~\cite{grewal2013caveat} proposes the notion of \emph{coercion
	evidence} and allows voters to cast multiple votes using the same credential.

\OpenProblem{
\begin{itemize}
\item Can we design usable, verifiable, coercion-resistant voting for a remote setting?
\end{itemize}
}

%% file: DoS.tex
\pagebreak
\subsection{Availability}
\emph{Denial-of-Service} (DoS) is an ever-present threat to elections which can
be mitigated but never fully eliminated. A simple service outage can
disenfranchise voters, and the threat of attack from foreign state-level
adversaries is a pressing concern.  Indeed, one of the countries that regularly
uses Internet voting, Estonia, has been subject to malicious
outages~\cite{estoniaattack}. 


A variant of DoS specific to the context of elections is \emph{selective DoS},
which presents a fundamentally different threat than general DoS. Voting
populations are rarely homogeneous, and disruption of service, for instance, in
urban (or rural) areas can skew results and potentially change election
outcomes. If DoS cannot be entirely eliminated, can service standards be
prescribed so that if an outcome falls below the standards it is vacated?
Should these standards be dependent on the reported margin of victory? What, if
any, recovery methods are possible?  Because elections are more vulnerable to
minor perturbations than most other settings, selective DoS is a concern which
cannot be ignored.

%% file: OtherRequirements.tex
\subsection{Usability}
\label{sec:Usability}
A voting system must be \emph{usable} by voters, poll-workers, election
officials, observers, and so on.  Voters who may not be computer literate---and
sometimes not literate at all---should be able to vote with very low error
rates.  Although some error is regarded as inevitable, it is also critical that
the interface not drive errors in a particular direction. For instance, a list
of candidates that crosses a page boundary could cause the
candidates on the second page to be missed. Whatever security mechanisms we add
to the voting process should operate without degrading usability, otherwise the
resulting system will likely be unacceptable. A full treatment of usability in
voting is beyond the scope of this paper. However, we note that E2E-V systems
(and I-voting systems, even when not E2E-V) add additional processes for voters
and poll workers to follow.  If verification processes can't be used properly by
real voters, the outcome will not be properly verified.  One great advantage of 
statistical audits is to shift complexity from voters to auditors.


\OpenProblems{
\begin{itemize}
\item How effectively can usability be integrated into the design process of a voting system?
\item How can we ensure full E2E-V, coercion resistance, etc., in a usable fashion?
\end{itemize}
}


\medskip
\subsection{Local Regulatory Requirements}

A variety of other mechanical requirements are often imposed by legal
requirements that vary among jurisdictions.  For example:
\begin{itemize}
    \setlength\itemsep{.5pt}
\item Allowing voters to ``write-in'' vote for a candidate
  not listed on the ballot.
\item Mandating the use of paper ballots (in some states without
  unique identifying marks or serial numbers; in other states {\em requiring} such marks)
\item Mandating the use of certain social choice functions (see~\ref{sec:complexVotingSchemes} Complex Election Methods below).
\item Supporting absentee voting.
\item Requiring or forbidding that ``ballot rotation'' be used (listing the candidates in different
  orders in different jurisdictions).
\item Requiring that voting equipment be certified under government guidelines.
  \end{itemize}

Newer electronic and I-voting systems raise important policy challenges for
real-world adoption. For example, in STAR-Vote~\cite{starvote-jets}, there will be
multiple copies of every vote record: mostly electronic records, but also paper
records. There may be instances where one is damaged or destroyed and the other
is all that remains. When laws speak to retention of ``the ballot'', that term
is no longer well-defined. Such requirements may need to be adapted to newer voting systems.

\medskip

\subsubsection{Complex Election Methods}
\label{sec:complexVotingSchemes}

Many countries allow voters to \emph{select, score, or rank}
candidates or parties.  Votes can then be tallied in a variety of complex
ways~\cite{brams2008mathematics,saari2012geometry}.  None of the requirements
for privacy, coercion-resistance, or the provision of verifiable evidence
change.  However, many tools that achieve these properties for traditional
"first-past-the-post" elections need to be redesigned.

An election method might be complex at the voting or the tallying end.  For
example, party-list methods such as D'Hondt and \StL{} have simple voting, in
which voters select their candidate or party, but complex proportional seat
allocation.  Borda, Range Voting, and Approval Voting allow votes to be quite
expressive but are simple to tally by addition.  Condorcet's method and related
functions~\cite{schulze2011new,tideman1987independence} can be arbitrarily
complex, as they can combine with any social choice function.  Instant Runoff
Voting (IRV) and the Single Transferable Vote (STV) are both expressive and
complicated to tally.  This makes for several challenges.

\OpenProblem {
\begin{itemize}
    \item Which methods for cast-as-intended verification 
        (e.g.\@ code voting~\cite{cha01:e-vote}) work for complex voting schemes? 
        \item How can we apply Risk-limiting audits to complex schemes? See Section~\ref{sec:complex-audits} for more detail.
    \item How can failures of the \emph{short ballot assumption}~\cite{rivest2007three} be mitigated with complex ballots?
    \item Can we achieve everlasting privacy for complex elections? 
\end{itemize}
}

%% file: TechTools.tex
\section{How can we secure voting?}
\label{sec:Tools}

\epigraph{These truths are self-evident but not self-enforcing}{\textit{ Barack Obama}}


The goal of this section and the next is to provide a state-of-the-art picture
of current solutions to voting problems and ongoing voting research, to
motivate further work on open problems, and to define clear directions both in
research and election policy.


\medskip
\subsection{The Role of Paper and Ceremonies}
\label{sec:paper-and-ceremonies}

Following security problems with direct-recording electronic voting systems
(DREs)~\cite{EVERESTmain,California-TTBR-site,feldman07diebold,wallach2006security},
many parts of the USA returned to the use of paper ballots. If secure custody of
the paper ballots is assumed, paper provides durable {\em evidence} required to
determine the correctness of the election outcome.  For this reason, when
humans vote from untrusted computers, cryptographic voting system
specifications often use paper for security, included in the notions of
dispute-freeness, dispute resolution, collection accountability  and
accountability \cite{DBLP:conf/ccs/KustersTV10} (all as defined in
Section~\ref{sec:trust}).


Note that the standard approach to dispute resolution, based on
non-repudiation, cannot be applied to the voting problem in the standard
fashion, because the human voter does not have the ability to check digital
signatures or digitally sign the vote (or other messages that may be
part of the protocol) unassisted.

Dispute-freeness or accountability are often achieved in a polling place through the use of cast
paper ballots, and the evidence of their chain of custody (e.g., wet-ink
signatures).
%
%
%
Paper provides an interface for data entry for the voter---not simply to enter
the vote, but also to enter other messages that the protocol might
require---and data on unforgeable paper serves many of the purposes of
digitally signed data. Thus, for example, when a voter marks a {\em Pr\^{e}t
\`{a} Voter}~\cite{RyanBHSX09} or {\em Scantegrity}~\cite{chaum08:scantegrity}
ballot, she is providing an instruction that the voting system cannot
pretend was something else. The resulting vote encryption has been physically
committed to by the voting system---by the mere act of printing the
ballot---before the voter ``casts'' her vote.


Physical ceremony, such as can be witnessed while the election is ongoing, also
supports verifiable cryptographic election protocols (see 
Section~\ref{subsubsec:cast-as-intended}).  Such ceremonies include
the verification of voter credentials, any generation of randomness if required
for the choice between cast and audit, any vote-encryption-verification
performed by election officials, etc.

The key aspect of these ceremonies is the chance for observers to see
that they are properly conducted.

\OpenProblem{
\begin{itemize}
\item Can we achieve dispute-resolution or -freeness without the use of paper
      and physical ceremony?
\end{itemize}
}


\subsection{Statistics and Auditing}
\label{sec:Statistics}

\input{Statistics}

\subsection{Cryptographic Tools and Designs}
\label{sec:Crypto}

\vspace{-5pt}
\subsubsection{Major Approaches to Voting Cryptography}

Typically E2E-V involves providing each voter with a \emph{protected
receipt}---an encrypted or encoded version of their vote---at the time the vote
is cast. 
The voter can later use her receipt to check whether her vote is included
correctly in the tabulation process.  Furthermore, given the set of encrypted
votes (as well as other relevant information, like the public keys), the
tabulation is \emph{universally verifiable}: anyone can check whether it is
correct.  To achieve this, most E2E-V systems rely on a public bulletin board,
where the set of encrypted ballots is published in an append-only fashion. 

The votes can then be turned into a tally in one of two main ways.
\emph{Homomorphic encryption}
schemes~\cite{DBLP:conf/focs/CohenF85,Benaloh:1987:VSE:914093} allow the tally
to be produced on encrypted votes.  \emph{Verifiable shuffling} transforms a
list of encrypted votes into a shuffled list that can be decrypted without the
input votes being linked to the (decrypted) output.  There are efficient ways
to prove that the input list exactly matches the
output~\cite{sako95:e-vote,neff01:e-vote,terelius2010proofs,bayer2012efficient,groth2010verifiable}. 
\looseness=-1


\vspace{-5pt}
\subsubsection{Techniques for Cast-as-Intended Verification}
\label{subsubsec:cast-as-intended}
How can a voter verify that her cast vote is the one she wanted?
\emph{Code Voting}, first introduced by Chaum~\cite{cha01:e-vote},  gives each voter a 
sheet of codes for each candidate.  Assuming the code sheet is valid, the 
voter can cast a vote on an untrusted machine by entering the code
corresponding to her chosen candidate and waiting to receive the correct
confirmation code.  Modern interpretations of code voting
include~\cite{DBLP:conf/acns/ZagorskiCCCEV13,joaquim2009veryvote,DBLP:conf/uss/Ryan11}.

Code voting only provides assurance that the correct voting code reached the
server, it does not of itself provide any guarantees that the code will
subsequently be correctly counted. A scheme that improves on this is Pretty
Good Democracy~\cite{RT09:PGD}, where knowledge of the codes is threshold shared in
such a way that receipt of the correct confirmation code provides assurance
that the voting code has been registered on the bulletin board by a threshold
set of trustees, and hence subsequently counted.




The alternative is to ask the machine to encrypt a vote directly, but verify
that it does so correctly.  Benaloh~\cite{BenalohSimple06} developed a simple
protocol to enable vote encryption on an untrusted voting machine. 
A voter uses a voting machine to encrypt any number of votes,
and casts only one of these encrypted votes. All
the other votes may be ``audited'' by the voter. If the encryption is audited,
the voting system provides a proof that it encrypted the vote correctly, and
the proof is public. The corresponding ballot cannot be cast as the
correspondence between the encryption and the ballot is now public, and the vote
is no longer secret. Voters take home receipts corresponding to the encryptions
of their cast ballots as well as any ballots that are to be audited.
They may check the presence of these
on a bulletin board, and the correctness proofs of the audited encryptions
using software obtained from any of several sources.
However, even the most dilligent voters need only check that their receipts
match the public record and that any ballots selected for audit display
correct candidate selections.  The correctness proofs are part of the public
record that can be verified by any individual or observer that is verifying
correct tallying.

\subsubsection{Formal models and security analyses of cast-as-intended verification protocols}
In addition to the work of Adida on assisted-human interactive proofs (AHIPs,
see~\cite{adida206:e-vote}), there has been some work on a rigorous
understanding of one or more properties of single protocols, including the work
of Moran and Naor~\cite{moran2010split,Moran-2010-Basing} and 
K\"{u}sters et al.~\cite{DBLP:conf/ccs/KustersTV10}.

There have also been formalizations of voting protocols with human
participants, such as by Moran and Naor~\cite{Moran-2010-Basing} (for a polling
protocol using tamper-evident seals on envelopes) and Kiayias {\it et al.}
\cite{kiayias2015ceremonies}.  However, there is no one model that is
sufficient for the rigorous understanding of the prominent protocols
used/proposed for use in real elections. The absence of proofs has led to the
overlooking of vulnerabilities in the protocols in the past,
see~\cite{karlof05:e-vote,khazaei2013randomized,DBLP:conf/wote/KelseRMC10,DBLP:conf/etrics/GogolewskiKKKLZ06}.

Many systems use a combination of paper, cryptography, and auditing to
achieve E2E-V in the polling place, including Markpledge~\cite{MarkPledge,adida2008efficient}, Wombat~\cite{Wombat,wombat2},
Demos~\cite{kiayias2015end}, \PaV~\cite{RyanBHSX09}, STAR-Vote~\cite{starvote-jets}, and Moran and Naor's
scheme~\cite{Moran-2006-receipt}.  Their properties are summarised more
thoroughly in the following section. 

The cryptographic literature has numerous constructions of end-to-end
verifiable election schemes (e.g.,~\cite{Fisher-2006-punchscan,Popoveniuc-2006-punchscan,RyanBHSX09,Carback-2010-ScantegrityII,rivest2007three,MarkPledge,Wombat,sandler08votebox,starvote-jets,joaquim2009veryvote}).  There are also
detailed descriptions of what it means to verify the correctness of the
output of E2E-V systems (e.g.,~\cite{kiayias2015end,Benaloh-1994-receipt,Moran-2006-receipt}).  
Others have attempted to define alternative forms of the
E2E-V properties~\cite{PopoveniucPerformance,cortierverifiability,DBLP:conf/ccs/KustersTV10}. 
There are also less technical explanations of
E2E-V intended for voters and election
officials~\cite{benaloh2015end,OVFE2EVIV}.
\filbreak

\OpenProblem{
    \vspace{-5pt}
 \begin{itemize}
    \item Can we develop a rigorous model for humans and the
          use of paper and ceremonies in cryptographic voting protocols?
    \item Can we rigorously examine the combination of statistical and cryptographic
          methods for election verification?
 \end{itemize}

    \vspace{-5pt}
}


\subsubsection{Techniques for Coercion Resistance}
\label{subsubsec:incoercibility-techniques}


Some simple approaches to coercion resistance have been suggested in
the literature.  These include allowing multiple votes with only the
last counting and allowing in-person voting to override remotely cast votes
(both used in Estonian, Norwegian, and Utah elections~\cite{estonia,norway,utah}). 
It is not clear that this mitigates coercion at all.  Alarm codes can also be
provided to voters: seemingly real but actually fake election credentials,
along with the ability for voters to create their own fake credentials.  Any
such approach can be considered a partial solution at best, particularly given the
usability challenges.

One voting system, {\em Civitas}~\cite{civitas}, based on a protocol by
Juels, Catalano and Jakobsson~\cite{juels05:e-vote}, allows voters to
vote with fake credentials to lead the coercive adversary into
believing the desired vote was cast. Note that the protocol must
enable universal verification of the tally from a list of votes cast
with both genuine and fake credentials, proving to the verifier that
only the ones with genuine credentials were tallied, without
identifying which ones they were.

\OpenProblem{
\begin{itemize}
    \item Can we develop cryptographic techniques that provide fully coercion
          resistant remote voting?
\end{itemize}
}

\subsubsection{Cryptographic Solutions in Complex Elections}

\label{sec:cryptographic-solutions}
Cast-as-intended verification based on creating and then challenging a vote
works regardless of the scheme ({\it e.g.} Benaloh challenges).  Cut-and-choose
based schemes such as \PaV{} and Scantegrity II need to be modified to work.


Both uses of end-to-end verifiable voting schemes in government
elections, the Takoma Park run of Scantegrity II and the
Victorian run of \PaV{}, used IRV (and one used STV). 
Verifiable IRV/STV counting that doesn't expose individual votes to the Italian
attack has been considered~\cite{shuffle-sum}, but may not be efficient enough
for use in large elections in practice, and was not employed in either
practical implementation.
\looseness=-1

\OpenProblems{
\begin{itemize}
    \item Is usable cast-as-intended
          verification for complex voting methods possible?
\end{itemize}
}

\subsubsection{Blockchains as a Cryptographic Solution}
Blockchains provide an unexpectedly effective answer to a long-standing problem
in computer science---how to form a consistent public ledger in a dynamic and
fully distributed environment in which there is no leader and participants may
join and leave at any time~\cite{nakamoto2008bitcoin}.  In fact, the blockchain
process effectively selects a "random" leader at each step to move things
forward, so this seems at first to be a natural fit for elections---citizens
post their preferences onto a blockchain and everyone can see and agree upon
the outcome of the election.  

However, blockchains and elections differ in significant ways. Elections
typically already have central authorities to play the leadership role, an
entity that administrates the election: what will be voted on, when,  who is
allowed to vote, etc.).  This authority can also be tasked with publishing a
public ledger of events.  Note that (as with blockchains) there need be no
special trust in a central authority as these tasks are all publicly
observable.  So to begin with, by simply posting something on a (digitally
signed) web page, an election office can do in a single step what blockchains
do with a cumbersome protocol involving huge amounts of computation.

Blockchains are inherently unaccountable.  Blockchain miners are individually
free to include or reject any transactions they desire---this is considered a
feature. To function properly in elections, a blockchain needs a mechanism to
ensure all legitimate votes are included in the ledger, which leads to another
problem: there's also no certainty in traditional blockchain schemes.  Disputes
are typically resolved with a "longest chain wins" rule.  Miners may have
inconsistent views of the contents of blockchains, but the incentives are
structured so that the less widely held views eventually fade away---usually.
This lack of certainty is not a desirable property in elections. 

In addition to lacking certainty and accountability, blockchains also lack
anonymity. While modifications can be made to blockchain protocols to add
anonymity, certainty, and accountability,  balancing these modifications on top
of the additional constraints of voting is difficult, and simpler
solutions already exist as we discuss.  

In short, blockchains do not address any of the fundamental problems in
elections, and their use actually makes things worse.

%% file: Statistics.tex


Two types of Risk Limiting Audits have been devised: \emph{ballot polling} and
\emph{comparison} \cite{LindemanStYa12,SOBA11,stark2008conservative}.  Both
types continue to examine random samples of ballots until either there is
strong statistical evidence that the outcome is correct, or until there has
been a complete manual tally.  ``Strong statistical evidence'' means that the
$p$-value of the hypothesis that the outcome is incorrect is at most $\alpha$,
within tolerable risk.  

Both methods rely on the existence of a \emph{ballot
manifest} that describes how the audit trail is stored.  Selecting the random
sample can include a public ceremony in which observers contribute by rolling
dice to seed a PRNG~\cite{corderoEtal06}.

\emph{Ballot-polling audits} examine random samples of individual ballots.  They
demand almost nothing of the voting technology other than the reported outcome.
When the reported outcome is correct, the expected number of ballots a
ballot-polling audit inspects is approximately quadratic in the reciprocal of
the (true) margin of victory, resulting in large expected sample sizes for small margins.

\emph{Comparison audits} compare reported results for randomly selected subsets of
ballots to manual tallies of those ballots.  Comparison audits require the
voting system to commit to tallies of subsets of ballots (``clusters'')
corresponding to identifiable physical subsets of the audit trail.  Comparison
audits have two parts: confirm that the outcome computed from the commitment
matches the reported outcome, and check the accuracy of randomly selected
clusters by manually inspecting the corresponding subsets of the audit trail.
When the reported cluster tallies are correct, the number of clusters a
comparison audit inspects is approximately linear in the reciprocal of the
reported margin.  The efficiency of comparison audits also depends
approximately linearly on the size of the clusters.  Efficiency is highest for
clusters consisting of individual ballots: individual cast vote records.  To
audit at the level of individual ballots requires the voting system to commit
to the interpretation of each ballot in a way that is linked to the
corresponding element of the audit trail.

In addition to RLAs, auditing methods have been proposed with
Bayesian~\cite{RivestShen-bayes} or
heuristic~\cite{Rivest-2015-diffsum} justifications.

All post-election audits implicitly assume that the audit trail is adequately
complete and accurate that a full manual count would reflect the correct
contest outcome.  \emph{Compliance audits} are designed to determine whether
there is convincing evidence that the audit trail was curated well, by checking
ballot accounting, registration records, pollbooks, election procedures,
physical security of the audit trail, chain of custody logs, and so on.
\emph{Evidence-based
elections}~\cite{StarkWagner-2012-Evidence-Based-Elections} combine compliance
audits and risk-limiting audits to determine whether the audit trail is
adequately accurate, and if so, whether the reported outcome is correct.  If
there is not convincing evidence that the audit trail is adequately accurate
and complete, there cannot be convincing evidence that the outcome is correct.
\subsubsection{Audits in Complex Elections}
\label{sec:complex-audits}

Generally, in traditional and complex elections, whenever an election margin is
known and the infrastructure for a comparison audit is
available, it is possible to conduct a rigorous risk-limiting comparison audit.
This motivates many works on practical margin computation for
IRV~\cite{Magrino:irv,Cary:irv,sarwate2013risk,blom2015efficient}.

However, such an audit for a complex election may not be efficient, 
which motivates the extension of Stark's \emph{sharper discrepancy measure} to
D'Hondt and related schemes~\cite{stark2014verifiable}.  For Schulze and some
related schemes, neither efficient margin computation nor any other form of RLA
is known (see \cite{hemaspaandra2013schulze}); a Bayesian
audit~\cite{RivestShen-bayes,chilingirian2016auditing} may nonetheless be used when one is able to
specify suitable priors.  





\OpenProblems{
\begin{itemize}
\item Can comparison audits for complex ballots be performed without exposing
    voters to ``Italian'' attacks?
\item Can RLAs or other sound statistical audits be developed for
    systems too complex to compute margins efficiently?
\item Can the notion of RLAs be extended to situations where physical
        evidence is not available (i.e. Internet voting)?
\end{itemize}
}


%% file: Designs.tex
\section{Current Solutions}
\label{sec:Designs}

\vspace{-5pt}

\epigraph{I am committed to helping Ohio deliver its electoral votes to the president next year.}
{\textit{Walden O'Dell, Diebold CEO, 2003}}


Below we provide a brief analysis of several real-world voting systems
developed by the scientific community. These systems use the properties
discussed in Sections \ref{sec:Requirements} and \ref{sec:Tools}. We include
both pollsite and remote systems. This collection is by no means exhaustive,
but hopefully the abundance of verifiable, evidence-based voting systems will
convince the reader that  there are significant technological improvements that
can greatly improve election security.  Our analysis is graphically represented
in Table 1.

\input{Table}

\subsection{Pollsite Systems}

The systems below were developed specifically with the requirements from
Section~\ref{sec:Requirements} in mind. As such, all satisfy the end-to-end
verifiability criteria from Section~\ref{sec:E2E}, and to a varying degree
provide
collection accountability, receipt-freeness, and coercion resistance.


\vspace{-2pt}
\subsubsection{\PaV{}}
\begin{figure}[h]
  \centering
  \includegraphics[width=0.5\linewidth]{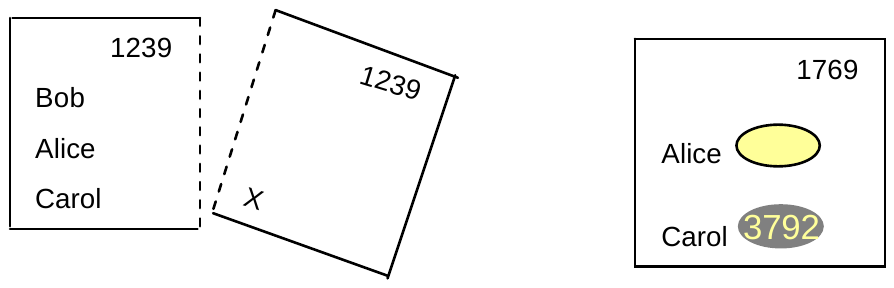}\hspace{18pt}
  \includegraphics[width=0.3\linewidth]{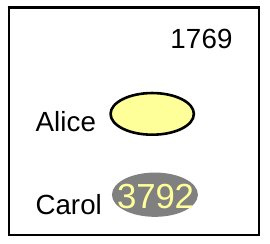}
  \caption{Marked ballots in Pr\^et-\`a-Voter~\cite{RyanBHSX09}
            (\emph{left}) and Scantegrity~\cite{chaum08:scantegrity} (\emph{right}).}
  \label{fig:ballots}
\end{figure}

\label{sec:PaV}
\input{Pret}

\vspace{-2pt}
\subsubsection{Scantegrity}
\label{sec:Scantegrity}

The {\em
Scantegrity}~\cite{chaum08:scantegrity,DBLP:journals/tifs/ChaumCCEPRRSSV09}
voter marks ballots that are very similar to optical scan ballots, with a
single important difference. Each oval has printed on it, in invisible ink, a
confirmation code---the encryption corresponding to this vote choice. When
voters fill the oval with a special pen, the confirmation number becomes
visible. The same functionality can be achieved through the use of scratch-off
surfaces.  


{\em Scantegrity II} was used by the City of Takoma Park for its municipal
elections in 2009 and 2011 \cite{Carback-2010-ScantegrityII},
the first secret-ballot election for public office known to use an E2E voting system
within the U.S.


\vspace{-2pt}
\subsubsection{VeriScan}
\label{sec:VeriScan}
VeriScan~\cite{VeriScan}, like Scantegrity, uses optical scan ballots.
But the ballots are ordinary -- using regular ink -- and are filled by voters
using ordinary pens.  Optical scanners used by VeriScan are augmented to hold
the ballot deposited by a voter and to print a receipt consisting of an
encryption of the selections made by the voter (or a hash thereof).

Once the receipt has been given to the voter by the scanner, the voter can
instruct the scanner to either retain the ballot or to return the ballot to
the voter.  A returned ballot should be automatically marked as no longer
suitable for casting and effectively becomes a challenge ballot as in STAR-Vote (below).

All encrypted ballots -- whether cast or retained by a voter -- are posted to
a public web page where they can be checked against voter receipts.  The cast ballots
are listed only in encrypted form, but the retained ballots are listed in both
encrypted and decrypted form so that voters can check the decryptions against their
own copies of the ballots.

\vspace{-2pt}
\subsubsection{STAR-Vote}
\label{sec:Star-Vote}
\input{STAR-Vote}


\vspace{-2pt}
\subsubsection{PPAT}
\label{sec:PPAT}
While many of the above schemes provide most of the required properties laid
out in Section \ref{sec:Requirements}, most do not account for everlasting
privacy. However, by integrating the Perfectly Private Audit Trail (PPAT)
\cite{cuvelier2013election}, many of the previously discussed systems can
attain everlasting privacy. Notably, PPAT can be implemented both with mixnet
schemes like Scantegrity \cite{chaum08:scantegrity} and Helios~\cite{adida08:e-vote} as
well as with homomorphic schemes like that used in
STAR-Vote~\cite{starvote-jets}.
\looseness=-1

\subsection{Remote Systems}

\subsubsection{Remotegrity}
\label{sec:Remotegrity}
The {\em Remotegrity}
\cite{DBLP:conf/acns/ZagorskiCCCEV13} voting system specification provides a
layer over local coded voting systems specifications to enable their use in a
remote setting. It is the only known specification that enables the voter to
detect and prove attempts by adversaries to change the remote vote.

Voters are mailed a package containing a coded-vote ballot and a credential
sheet. The sheet contains authorization codes and lock-in codes under
scratch-offs, and a return code. To vote, voters scratch-off an authorization
code at random and use it as a credential to enter the candidate code. The
election website displays the entered information and the return code, which
indicates to the voter that the vote was received.  If the website displays the
correct information, the voter locks it in with a random lock-in code. If not,
the voter uses another computer to vote, scratching-off another authorization
code. For voter-verifiability, voters may receive multiple ballots, one of
which is voted on, and the others audited.

The credential authority (an insider adversary) can use the credentials to vote
instead of the voter. If this happens, the voter can show the
unscratched-off surface to prove the existence of a problem. Remotegrity thus
achieves E2E-V, collection accountability, and software independence.
Since there is no secret ballot guarantee, there is no coercion resistance.

{\em Remotegrity} was made available to absentee voters in the 2011
election of the City of Takoma Park, alongside in-person voting provided by {\em Scantegrity}.

\subsubsection{Helios}
\label{sec:Helios}
\input{helios}

\subsubsection{Selene}
\label{sec:Selene}
\input{Selene}

\OpenProblems{
\begin{itemize}
\item Is there a cast-as-intended method that voters can
    execute successfully without instructions from pollworkers?
\item Is it possible to make E2E-V  protocols simpler for election officials
and pollworkers to understand and administer?

\end{itemize}
}

%% file: Table.tex
\begin{table*}
\centering

\newcolumntype{R}[2]{%
    >{\adjustbox{angle=#1,lap=\width-(#2)}\bgroup}%
    l%
    <{\egroup}%
}

\newcommand*\rot{\multicolumn{1}{R{60}{1em}}}

\newcommand{\YES}{\CIRCLE}
\newcommand{\NO}{\Circle}
\newcommand{\PARTLY}{\LEFTcircle}
\newcommand{\DASH}{\hskip1.33pt--}

\newcommand{\SmallTrials}{1}
\newcommand{\Pending}{2}
\newcommand{\PrivateSector}{3}
\newcommand{\AbsenteeOnly}{4}
\newcommand{\Revote}{5}
\newcommand{\PPATS}{6}
\newcommand{\Audit}{7}
\newcommand{\Email}{8}
\newcommand{\Temporary}{9}
\newcommand{\Query}{10}
\newcommand{\CodeSheets}{11}
\newcommand{\Assurance}{12}
\newcommand{\Active}{13}
\newcommand{\Unforgeable}{14}


\centering
\begin{threeparttable}

\begin{tabular}{rlllllllllllll}

    \relax &

    \rot{fielded} &
    \rot{coercion resistance} &
    \rot{everlasting privacy} &

    \rot{software independence} &

    \rot{take-home evidence} &
    
    \rot{ballot cast assurance} &
    \rot{collection accountable} &

    \rot{verifiably cast-as-intended} &
    \rot{verifiably collected-as-cast} &
    \rot{verifiable counted-as-collected} &

    \rot{paper/electronic/hybrid} &
    \rot{write-ins supported} &
    \rot{preferential ballots supported}

    \\
 \toprule

\multicolumn{1}{l}{\bf Poll-site techniques in widespread use}\smallskip \\

Hand-counted in-person paper & \YES & \YES & \YES & \YES & \NO & \NO & \NO & \YES & \NO & \YES\tnote{\Audit} & p & \YES & \YES  \\
Optical-scan in-person paper & \YES & \YES & \YES & \YES & \NO & \NO & \NO  & \YES & \NO & \YES\tnote{\Audit}& h & \YES & \YES  \\
DRE (with paper audit trail) & \YES & \YES & \NO & \YES & \NO  & \NO & \NO & \PARTLY\tnote{\Audit} & \YES & \YES\tnote{\Audit} & h & \YES & \YES \\
Paperless DRE                & \YES & \YES & \NO & \NO  & \NO  & \NO & \NO & \NO & \NO & \NO & e & \YES & \YES \\

\midrule

\multicolumn{1}{l}{\bf Poll-site systems from research}\smallskip \\

Pr\^et-\`a-voter~\cite{RyanBHSX09} & \YES\tnote{\SmallTrials} & \YES & \NO & \YES & \YES & \PARTLY\tnote{\Assurance} & \YES & \YES & \YES & \YES & h & \NO & \YES  \\
Scantegrity~\cite{chaum08:scantegrity} & \YES\tnote{\SmallTrials} & \YES & \NO & \YES & \YES & \PARTLY\tnote{\Assurance} & \YES & \YES & \YES & \YES & h & \PARTLY & \PARTLY \\
 STAR-Vote~\cite{starvote-jets} & \PARTLY\tnote{\Pending} & \YES & \PARTLY\tnote{\PPATS} & \YES & \YES & \PARTLY\tnote{\Active} & \PARTLY\tnote{\Unforgeable} & \YES & \YES & \YES & h & \NO & \NO  \\

Wombat~\cite{Wombat} & \PARTLY\tnote{\PrivateSector} & \YES & \PARTLY\tnote{\PPATS} & \YES & \YES & \PARTLY\tnote{\Active} & \PARTLY\tnote{\Unforgeable} & \YES & \YES & \YES & h & \NO & \NO  \\
VeriScan~\cite{VeriScan} & \NO & \YES & \PARTLY\tnote{\PPATS} & \YES & \YES & \PARTLY\tnote{\Active}  & \PARTLY\tnote{\Unforgeable} & \YES & \YES & \YES & h & \NO & \PARTLY \\
Scratch and Vote~\cite{adida06:e-vote} & \NO & \YES & \NO & \YES & \YES & \PARTLY\tnote{\Assurance} & \YES & \YES & \YES & \YES & h & \PARTLY & \PARTLY  \\
MarkPledge~\cite{MarkPledge} & \NO & \YES & \PARTLY\tnote{\PPATS} & \YES & \YES & \YES & \YES & \YES & \YES & \YES & e & \NO & \NO  \\
ThreeBallot~\cite{rivest06:e-vote}& \NO & \YES & \YES & \YES & \YES & \PARTLY\tnote{\Assurance} & \YES & \YES & \PARTLY & \YES & h & \NO & \NO \\

\midrule

\multicolumn{1}{l}{\bf Remote voting systems and techniques}\smallskip \\

Helios~\cite{adida08:e-vote} & \PARTLY\tnote{\PrivateSector} & \NO\tnote{\Revote} & \PARTLY\tnote{\PPATS} & \YES & \YES
\tnote{\Email} & \PARTLY & \PARTLY\tnote{\Audit} & \YES & \YES & \YES & e & \NO & \NO  \\
Remotegrity~\cite{DBLP:conf/acns/ZagorskiCCCEV13} & \PARTLY\tnote{\SmallTrials} & \NO & \PARTLY\tnote{\PPATS} & \YES & \YES & \NO & \YES & \YES & \YES & \YES & h & \NO & \PARTLY \\
Civitas~\cite{civitas} & \PARTLY\tnote{\PrivateSector} & \YES & \PARTLY\tnote{\PPATS} & \NO & \YES\tnote{\Email} & \NO & \PARTLY & \YES & \YES & \YES & e & \PARTLY & \YES  \\
Selene~\cite{selene15} & \NO & \YES & \PARTLY\tnote{\PPATS} & \YES & \NO & \PARTLY\tnote{\Audit} & \YES & \YES & \YES & \YES & e & \NO & \YES \\
Norway~\cite{norway} & \YES & \NO\tnote{\Revote} & \NO & \NO & \NO & \NO & \PARTLY\tnote{\CodeSheets} & \NO & \NO & \PARTLY\tnote{\Audit} & e & \YES & \YES   \\
Estonia~\cite{estonia} & \YES & \NO\tnote{\Revote} & \NO & \NO & \PARTLY\tnote{\Temporary} & \NO & \NO & \NO & \NO & \NO & e & \NO & \NO  \\
iVote~\cite{halderman2015new} & \YES\tnote{\AbsenteeOnly} & \NO\tnote{\Revote} & \NO & \NO & \NO & \NO & \NO & \PARTLY\tnote{\Query} & \PARTLY\tnote{\Query} & \NO & e & \NO & \YES \smallskip\\

Paper ballots returned by postal mail & \YES & \NO & \YES & \YES & \NO & \NO & \NO & \NO & \NO & \YES\tnote{\Audit} & p & \YES & \YES \\

\bottomrule
\end{tabular}

\begin{tablenotes}\footnotesize

    \centering
    \item {\footnotesize \YES ~= provides property\quad\NO ~= does not provide property\quad\PARTLY ~= provides property with provisions} \vspace{\baselineskip}\\

    \newcommand{\Tnote}[1]{\textsuperscript{#1}}

    \parbox{1in}{\begin{tabular}{@{}r@{\hspace{3pt}}l}
    \Tnote\SmallTrials & Used in small trial elections \\
    \Tnote\Pending & Pending deployment \\
    \Tnote\PrivateSector & Used in private sector elections \\
    \Tnote\AbsenteeOnly & Absentee voting only \\
    \Tnote\Revote & Allows multiple voting \\
    \end{tabular}}\enspace
    \parbox{1in}{\begin{tabular}{@{}r@{\hspace{3pt}}l}
    \Tnote\PPATS & Possible with PPAT \\
    \Tnote\Audit & With sufficient auditing \\
    \Tnote\Email & Receipts sent by email \\
    \Tnote\Temporary & Temporary email receipt \\
    \Tnote\Query & Queryable (phone system) \\
    \end{tabular}}\enspace
    \parbox{1in}{\begin{tabular}{@{}r@{\hspace{3pt}}l}
    \Tnote\CodeSheets & Queryable (code sheets) \\
    \Tnote\Assurance & \multirow{2}{1.2in}{Enhanced with pre- and post-election auditing} \\ \\ 
    \Tnote\Active & \multirow{2}{1.2in}{Enhanced with auditing during elections} \\ \\
    \Tnote\Unforgeable & \multirow{2}{1.2in}{To the extent the paper resists forgery} \\ \\
    \end{tabular}}

\end{tablenotes}

\end{threeparttable}


\caption{\textbf{Applying our threat model to fielded and proposed voting schemes} --- 
Note that certain features like credentialing and availability are excluded, as these factors impact all
systems in roughly equivalent ways. The Utah system has not been made available for rigorous security analysis, and is
excluded. 
}

\label{tab:designs}
\end{table*}

%% file: Pret.tex
{\em Pr\^{e}t \`{a} Voter}~\cite{RyanBHSX09} ballots list the candidates in a
pseudo-random order, and the position of the voter's mark serves as an encryption of
the vote. The ballot also carries an encryption of the candidate ordering,
which can be used, with the mark position, to obtain the vote.  Voters can
audit ballots to check that the random candidate order they are shown matches
the encrypted values on their ballot.


\paragraph{vVote} In the 2014 state election the Australian state of Victoria
conducted a small trial of end-to-end verifiable pollsite voting, using a
system called vVote derived from \PaV{}~\cite{vVoteTISSEC}.


%% file: STAR-Vote.tex
STAR-Vote~\cite{starvote-jets} is an E2E-V, in-person voting system designed
jointly with Travis County (Austin), Texas, and is scheduled for wide-spread
deployment in 2018. STAR-Vote is a DRE-style touch-screen system, which
prints a human-readable paper ballot which is deposited into a ballot box. The
system also prints a receipt that can be taken home. These two printouts serve
as evidence for audits. 

STAR-Vote encodes a Benaloh-style cast-or-spoil
question~\cite{BenalohSimple06} as the depositing of the
ballot into the ballot box. Each voting machine must commit to the voter's ballot
without knowing if it will be deposited and counted or spoiled and thereby
challenged.  

STAR-Vote posts threshold encrypted cast and spoiled ballots to a web bulletin board. Voters can then
check that their cast ballots were included in the tally, or that the system
correctly recorded their vote by decrypting their challenged ballots. STAR-Vote
is collection accountable only to the extent that paper receipts and ballot summaries
are resistent to forgery.  It is coercion resistant and software
independent, and allows for audits of its paper records.


%% file: helios.tex
Helios~\cite{adida08:e-vote,HeliosElection} is an E2E-V Internet voting system.
Voters visit a web page ``voting booth'' to enter their selections.  After
voters review  their ballots, each ballot is encrypted using a threshold key
generated during election set up.

Voters cast a ballot by entering credentials supplied for this election.
Alternatively, voters can anonymously spoil their ballots to decrypt them, to
show that their selections were accurately recorded. Voters can cast multiple
ballots with only the last one retained, as a weak means of coercion mitigation.

When the election closes, the cast votes are verifiably tallied---either using
homomorphic tallying or a mixnet. Independent verifiers have
been written to check the tallying and decryptions of each spoiled
ballot. Confirmation that the vote is received is then emailed to the voter. 
Helios is used for elections by a variety of universities and professional societies
including the Association for Computing Machinery and the International Association
for Cryptologic Research.
Helios lacks collection accountability, but is still E2E-V and software
independent through its spoil function.

%% file: Selene.tex
Selene~\cite{selene15} is a remote E2E-V system that revisits the tracker
numbers of Scantegrity, but with novel cryptographic constructs to counter the
drawbacks. Voters are notified of their tracker after the vote/tracker pairs
have been posted to the web bulletin board, which allows coerced
voters to identify an alternative tracker pointing to the coercer's required
vote. Voter verification is much more
transparent and intuitive, and voters are not required to check the presence of
an encrypted receipt.  For the same reasons as Remotegrity, Selene is software
independent and provides collection accountability.

%% file: InternetVoting.tex
\section{Internet Voting}
\label{sec:InternetVoting}


\epigraph{``People of Dulsford,'' began Boris, ``I want to assure you that as
 your newly elected mayor I will not just represent the people who voted for me
 ...'' \\
 \vspace{5pt} ``That's good,'' said Derrick, ``because no-one voted for him.'' \\
 \vspace{5pt} ``But the people who didn't vote for me as well,'' said Boris. \\ 
 \vspace{5pt} There was a
 smattering of half-hearted clapping from the crowd.} {\textit{R. A. Spratt,}
 Nanny Piggins and the Race to Power}


In this section we present the challenges of secure Internet voting through a
set of (possibly contradictory) requirements. No system has addressed the
challenges sufficiently so far, and whether it is possible to do so remains an
open problem.  We begin by introducing prominent contemporary instances of
I-voting as case studies. Then we examine the Internet voting threat model,
along the way showing how these Internet systems have failed to adequately
defend themselves. We look at voter authentication, verification of the
correctness of a voting system's output, voter privacy and coercion resistance,
protections against denial-of-service, and finally the usability and regulatory
constraints faced by voting systems.

One major roadblock faced exclusively by I-voting is the underlying
infrastructure of the Internet. The primary security mechanism for Internet
communication is Transport Layer Security (TLS), which is constantly evolving
in response to vulnerabilities.
For instance, the website used in the iVote system  was vulnerable to the TLS
\textsc{FREAK}~\cite{dambh:freak} and LogJam~\cite{weakdh-ccs15}
vulnerabilities.  Researchers discovered this \emph{during} the election period
and demonstrated that they could exploit it to steal
votes~\cite{halderman2015new}. At the time, LogJam had not been publicly
disclosed, highlighting the risk to I-voting from zero-day vulnerabilities.
Internet voting systems \emph{must} find ways to rely on properties like
software independence and E2E-V before they can be considered trusted.

In 2015, the U.S. Vote Foundation issued an export report on the viability of
using E2E-verifiability for Internet voting~\cite{OVFE2EVIV}.
The first two conclusions of the report were as follows.
\begin{enumerate}
\item{Any public elections conducted over the Internet must be end-to-end verifiable.}
\item{No Internet voting system of any kind should be used for public elections before
end-to-end verifiable in-person voting systems have been widely deployed
and experience has been gained from their use.}
\end{enumerate}

Many of the possible attacks on I-voting systems could be performed on 
postal voting systems too.  The main difference is the likelihood that a 
very small number of people could automate the manipulation of a very large
number of votes, or a carefully chosen few important votes, without detection.

\input{InternetSystems}

\subsection{E2E-V I-voting in Government Elections}
Internet voting presents numerous challenges that have not been adequately addressed.
First among these is the coercion problem which is shared with other remote voting
systems in widespread use today (such as vote-by-mail).  However, I-voting exacerbates
the problem by making coercion and vote-selling a simple matter of a voter providing
credentials to another individual.

Client malware poses another significant obstacle.  While E2E-verifiability mitigates
the malware risks by providing voters with alternate means to ensure that their votes
have been properly recorded and counted, many voters will not avail themselves of these
capabilities.  We could therefore have a situation were a large-scale fraud is observed
by a relatively small number of voters.  While the detection of a small number of
instances of malfeasance can bring a halt to an election which provides
collection accountability, the required evidence can be far more fleeting and
difficult to validate in an Internet setting.  An election should not be overturned
by a small number of complaints if there is no substantive evidence to support
these complaints.

Targeted denial-of-service is another serious unresolved threat to I-voting.
Ordinary denial-of-service (DoS) is a common threat on the Internet, and means
have been deployed to mitigate --- although not eliminate --- these threats.
The unique aspect in elections is that while ordinary DoS can slow commerce
or block access to a web site for a period, the effects of a targeted DoS attack
on an election can be far more severe.  Since voting paterns are far from
homogeneous, an attacker can launch a targeted DoS attack against populations
and regions which are likely to favor a particular candidate or position.
By merely making it more difficult for people in targeted populations to vote,
the result of an election can be altered.  As yet, we have no effective
mitigations for such attacks.

Finally, as was observed in the U.S. Vote Foundation study~\cite{OVFE2EVIV},
we simply don't yet have much experience with large-scale deployments of
E2E-verifiable election systems in the simpler and more manageable setting
of in-person voting.  It would be angerous to jump directly to the far more
challenging setting of Internet voting with a heavy dependence on a technology
that has not previously been deployed at scale.




\subsection{Alternatives to Internet Voting}
There are numerous alternatives to Internet voting that can help enfranchise voters who can not
easily access a poll site on the day of an election.

Early voting is in widespread use throughout the U.S.  By extending the voting window
from a single day to as much as three weeks, voters who may be away or busy
on the date of an election can be afforded an opportunity to vote in person,
at their convenience, at a poll site with traditional safeguards.
Early voting also mitigates many of the risks of traditional systems since,
for example, an equipment failure ten days prior to the close of an election
is far less serious than one that takes place during a single day of voting.

Some U.S. jurisdictions have adopted a \emph{vote center} system in which voters
may vote in person outside of their home precincts.  This option has been
facilitated by the use of electronic poll books, and it allows voters to, for instance,
vote during a lunch break from work if they will be away from their homes during
voting hours.  The vote center model could potentially be extended from the current
model of voters away from their home precincts but still within their home counties
by allowing voters to use any poll site in the state or country.
It would even be possible to establish remote voting kiosks overseas in embassies,
conslates, or other official sites, and roming voting kiosks could be established
with as little as two poll workers and a laptop computer.  Security and accountability
in all of these non-local voting scenarios can be greatly enhanced by the use of
E2E-verifiability.

Blank-ballot electronic delivery is another option which has gained in popularity.
While there are numerous risks in using the Internet for casting of ballots,
the risks a far less in simply providing blank ballots to voters.
Electronic delivery of blank-ballots can save half of the round-trip time that
is typical in absentee voting, and traditional methods of ballot return can be
used which are less susceptible to the large-scale attacks that are possible
with full Internet voting.





%% file: InternetSystems.tex

\subsection{I-voting in Government Elections}

\paragraph{Estonia~\cite{Estonia2005}} Estonia's I-voting deployment---the
largest in the world by fraction of the electorate---was used to cast nearly a
third of all votes in recent national elections~\cite{kov13stats}.  The
Estonian system uses public key cryptography to provide a digital analog of the
``double envelope'' ballots often used for absentee voting~\cite{Estonia2005}.
It uses a national PKI system to authenticate voters, who encrypt and digitally
sign their votes via client-side software.  Voters can verify\footnote{We use
the term ``verify'' loosely in this subsection; these systems provide no
guarantee that what is shown when voters ``verify'' their votes proves anything
about the correctness of vote recording and processing. 
see~\ref{sec:trust}.  } that their votes were correctly received using a
smartphone app, but the tallying process is only protected by procedural
controls~\cite{ivotingverification:github}. The voting system does not provide
evidence of a correct tally, nor does it provide evidence that the vote was
correctly recorded if the client is dishonest. A 2013 study showed that the
Estonian system is vulnerable to vote manipulation by state-level attackers and
client-side malware, and reveals significant shortcomings in officials'
operational security~\cite{estonia}.

\paragraph{iVote~\cite{ivote-overview}}
The largest online voting trial by absolute number of votes occurred in 2015 in
New South Wales, Australia, using a web-based system called iVote.  It received
280,000 votes out of a total electorate of over 4~million.  The system included
a telephone-based vote verification service that allowed voters to dial in and
hear their votes read back in the clear. A limited server-side auditing
process was performed only by auditors selected by the electoral authority.
Thus no evidence was provided that received votes were correctly included in 
the tally. At election time, the electoral commission declared that, 
``1.7\% of electors who voted using iVote also used the verification service and none of them identified any anomalies with their vote.''
It emerged more than a year later that 10\% of verification attempts had failed to retrieve any
vote at all.  This error rate, extrapolated to all 280,000 votes, would have been
enough to change at least one seat.

\paragraph{Norway~\cite{norway}} In 2011 and 2013 Norway ran trials of an
I-voting system. In the 2013 trial, approximately 250,000 voters (7\% of the
Norwegian electorate) were able to submit ballots online~\cite{norway-carter}.
Voters are given precomputed encrypted return codes for the various candidates
they can vote for. Upon submitting a ballot, the voter receives an SMS message
with the return code computed for the voter's selections. In principle, if the
return codes were kept private by the election server, the voter knows the
server correctly received her vote. This also means that ballots must be
associated with the identity of those who cast them, enabling election
officials to possibly coerce or selectively deny service to voters. The voting
system did not provide publicly verifiable evidence of a correct tally. 

\paragraph{Switzerland~\cite{SwissFCReqs}} In Switzerland, the Federal
Chancellery has produced a clear set of requirements.  More stringent
verifiability properties come into force as a larger fraction of the votes are
carried over the Internet.  Many aspects of this way of proceeding are
admirable. However, the final systems are dependent on a code-verification
system, and hence integrity depends on the proper and secret printing of the
code sheets.  If the code-printing authorities collude with compromised
devices, the right verification codes can be returned when the votes are wrong.


\paragraph{Utah~\cite{utah}}
In March 2016 the Utah Republican party held its caucus, running pollsite
voting in addition to an online system. Voters could register through a
third-party website and have a voting credential sent to their phone via SMS or
email. Any registered voter could receive a credential, but as the site was
unauthenticated, anyone with a voter roll\footnote{That is, a publicly
available list of registered voters, their party affiliations, home addresses, and
other relevant information} could submit any registered voter's information and
receive that person's credential. On the day of the election, said credentials
were used to log onto the website \url{ivotingcenter.us} to fill out and submit
ballots. The system provided voters with a receipt code that voters could check on the election website. The system does not provide evidence that the vote was correctly recorded if the client is dishonest, nor does it provide evidence of a correct tally. Election day saw many voters fail to receive their voting credentials
or not be able to reach the website to vote at all, forcing as many as 13,000
of the 40,000 who attempted to register to vote online to either vote in person
or not vote at all~\cite{utah-caucus-article}.

\smallskip

All of these systems place significant trust in unverifiable processes,
at both client and server sides, leading to serious weaknesses in privacy and
integrity.  Their faults demonstrate the importance of a clear and careful
trust model that makes explicit who does and does not have power over the votes
of others, and reinforce the importance of providing convincing evidence of an
accurate election outcome. 

%% file: ALookAhead.tex
\section{A Look Ahead} \label{sec:ALookAhead}

\epigraph{There is no remedy now to a process that was so opaque that it could have been manipulated at any stage}{\textit{Michael Meyer-Resende and Mirjam Kunkler, on the Iranian 2009 Presidential election}}

Voting has always used available technology, whether pebbles dropped
in an urn or marked paper put in a ballot box; it now uses computers,
networks, and cryptography.  
The core requirement, 
to provide public evidence of the right result from secret ballots,
hasn't changed in 2500 years.

Computers can improve convenience and accessibility over plain paper and manual counting.
In the polling place there are good solutions, including Risk Limiting Audits and end-to-end verifiable systems. These must be more widely deployed and their options for verifying the election result must actually be used.

Many of the open problems described in this paper---usable and
accessible voting systems, dispute resolution, incoercibility---come
together in the challenge of a remote voting system that is verifiable
and usable without supervision. The open problem of a system
specification that (a) does not use any paper at all and (b) is based
on a simple procedure for voters and poll workers, will motivate
researchers for a long time.  Perhaps a better goal is a hybrid system combining paper evidence with some auditing or cryptographic verification. 



Research in voting brings together knowledge in many fields---cryptography,
systems security, statistics, usability and accessibility, software
verification, elections, law and policy to name a few---to address a critical
real-world problem.  

The peaceful transfer of power depends on confidence in the electoral process. That confidence should not automatically be given to any outcome that seems plausible---it must be earned by producing evidence that the election result is what the people chose.  
Insisting on evidence reduces the opportunities for fraud,
hence bringing greater
security to citizens the world over.